\documentclass[preprint,3p,authoryear]{elsarticle}
\usepackage[utf8]{inputenc}
\usepackage[T1]{fontenc}

\usepackage{amssymb}
\usepackage{amsmath}
\usepackage{commands}
\usepackage{url}

\usepackage{tabularx}

\usepackage{bigfoot} 

\usepackage{xcolor}
\definecolor{darkblue}{rgb}{0.15, 0.15, 0.45}
\definecolor{darkgreen}{rgb}{0.17, 0.4, 0.21}
\definecolor{urlcolor}{rgb}{.41,.25,.45}
\definecolor{gray75}{gray}{0.75}
\usepackage[colorlinks=true,
            linkcolor=darkblue,
            citecolor=darkblue,
            urlcolor=urlcolor,
	    pdftitle={ana_cont},
	    pdfauthor={Josef Kaufmann and Karsten Held}]{hyperref}
\usepackage[capitalize,nameinlink]{cleveref}

\journal{Computer Physics Communications}

\begin{document}

\begin{frontmatter}



\title{ana\_cont: Python package for analytic continuation}


\author{Josef Kaufmann and Karsten Held}

\affiliation{organization={Institute for Solid State Physics, TU Wien},
            addressline={Wiedener Hauptstrasse 8-10}, 
            city={1040 Vienna},
            country={Austria}}

\begin{abstract}
  We present the Python package \texttt{ana\_cont} for the analytic continuation of
  fermionic and bosonic many-body Green's functions by means of either the \pade\ approximants
  or the maximum entropy method.
  The determination of hyperparameters and the implementation are
  described in detail. The code is publicly available on GitHub,
  where also documentation and learning resources are provided.
\end{abstract}

%


\end{frontmatter}


\noindent
{\bf PROGRAM SUMMARY}

\begin{small}
\noindent
  {\em Program Title:} ana\_cont\\
  {\em Licensing provisions:} MIT\\
  {\em Operating system:} Linux, Unix \\
  {\em Available open source at:} \anaconturl, to be published also in the Mendeley Data repository\\
  {\em Programming language:} \verb=Python= \\
  {\em Required dependencies:} \verb=Python= $(\geq 3.6)$, \verb=numpy=, \verb=scipy=, \verb=matplotlib=, \verb=h5py=, \verb=PyQt5=, \verb=Cython=\\
  {\em Supplementary material:} Test case files, tutorials, and instructions\\
  {\em Nature of problem:} Analytic continuation of correlation functions from Matsubara frequencies/imaginary time to real frequencies.\\
  {\em Solution method:} \pade\ interpolation, maximum entropy method\\
  {\em Additional comments including restrictions and unusual features:} The most important features can be accessed
  through the graphical user interface. For more flexibility, it is recommended to use
  the code as a library and write problem specific scripts.\\

\end{small}


\section{Introduction}
One of the beauties of complex analysis is that if we know an analytic function of a complex variable, i.e. a holomorphic function,
on an (open) subdomain we know it on any connected domain. The many-body Green's functions of quantum field theory  \citep{Abrikosov} are such holomorphic functions, and often it is more convenient or also numerically more stable to do calculations for imaginary frequencies or times. This is possible because of the analytic continuation. But, it eventually requires an analytic continuation back to real (physical) frequencies and times  at the end. Take, for instance, the time propagation $e^{-i Ht}$ with the Hamiltonian $H$ from time zero to $t$ (Planck constant $\hbar\equiv 1$). An analytic continuation to complex times  $e^{-H\tau}$, here also called Wick rotation  \citep{Abrikosov} $t\rightarrow  -i\tau$,  allows  us to treat the time propagation on the same footing as the Boltzmann operator  $e^{-H\beta}$ ($\beta=1/T$: inverse temperature;  Boltzmann constant $k_B\equiv 1$).
Not only because of the unified time propagation, 
(semi)-analytical calculations are often easier to formulate in imaginary times or frequencies.

Numerical approaches such as plain-vanilla exact diagonalization or solving the parquet equations \citep{Bickers2004,victory} for imaginary Matsubara frequencies avoid poles on the real frequency axis and connected numerical instabilities or discretization errors.  Some numerical methods such as quantum Monte-Carlo simulations \citep{Gull2011a,Wallerberger2019} do not even work properly for real times.
In all of these cases one needs, at the very end, an analytic continuation back to the real axis if physically-relevant  dynamics is calculated.

Two state-of-the-art approaches to this end are the \pade\ approximation \citep{NumericalRecipes} and the maximum entropy (MaxEnt) method \citep{JarrellGubernatis1996}. For more recent advances cf.\  \citet{Bergeron2016}, \citet{Levy2017}, \citet{Kraberger2017}. Also alternatives such as sparse modeling of Green's functions in the intermediate representation \citep{SpM} or neural networks  \citep{Fournier2020} are discussed in the literature.

In this paper, we present a pedagogical introduction to the analytic continuation of Green's functions by means of the  \pade\ approximation \citep{NumericalRecipes} and the maximum entropy (\maxent) method. We exemplify  how noise and the choice of parameters affect the results. Last but not least, we introduce the OpenSource Python package \texttt{ana\_cont}, discuss its implementation and usage.

\paragraph*{Outline}
The manuscript is structured as follows: We first give a minimal introduction
to many-body Green's functions and their analytic properties in \cref{sec:analytic-properties}.
After that, we review the method of \pade\ interpolation  in \cref{sec:pade}.
Then we show how a probabilistic approach, using Bayes' theorem, leads to the
maximum entropy method (\maxent) in \cref{sec:maxent}.
In the subsequent \cref{sec:technical} we discuss technical aspects
of \maxent, which allows us to establish stable workflows for determining 
hyperparameters. 
Then, in \cref{sec:package}, we present our actual implementation.
In detail we describe numerical aspects, as well as how to use it
for self-energies and susceptibilities. 
The installation procedure is explained in \cref{sec:installation}.
Finally, we conclude our work in \cref{sec:conclusion}.

\section{Many-body Green's functions}\label{sec:analytic-properties}
\noindent
Many properties of a system of interacting electrons
are encoded in the retarded one-particle Green's function \citep{Abrikosov}
\begin{equation}
  \label{eq:g-ret-t}
  G_R(t) = -i \Theta(t) \big\langle \{ \cee(t), \cdag(0) \} \big\rangle.
\end{equation}
This definition uses second quantization, where an operator
$\cdag(t)$ creates an electron at time $t$, and $\cee(t)$ annihilates it.
For the sake of brevity we omit further indices such as momentum or site here, 
because for the analytic continuation only the general analytic properties are of relevance.
The Heaviside function $\Theta(t)$ is $0$ for all times $t<0$, and
$1$ for $t>0$. Angular brackets $\langle\cdots\rangle$ denote
averaging over a grand canonical ensemble.
It is usually more instructive to look at the retarded Green's function
in frequency domain,
\begin{equation}
  \label{eq:g-ret-w}
  G_R(\omega) = \int_{-\infty}^{\infty} dt\; e^{i\omega t}\, G_R(t),
\end{equation}
where frequencies $\omega$ are equivalent to energies by setting $\hbar\equiv 1$.

However, computational methods for many-body systems often employ the Matsubara
formalism \citep{Matsubara1955}, where correlation functions are calculated
on the imaginary time or frequency axis. 
The fermionic one-particle Green's function in imaginary time $\tau$ is
\begin{equation}
  \label{eq:g-imag-t}
  G(\tau) = -\big\langle \timeorder_\tau \cee(\tau) \cdag(0) \big\rangle
\end{equation}
with the time ordering operator $\timeorder_\tau$,
and by Fourier transform one obtains 
\begin{equation}
  \label{eq:g-mats}
  G(i\omega_n) = \int_{0}^{\beta} d\tau\; e^{i\omega_n \tau}\, G(\tau),
\end{equation}
where by $i\omega_n$ we denote fermionic Matsubara frequencies
$\omega_n = (2n + 1)\pi/\beta$ for fermions.

\begin{figure}
  \centering
  \includegraphics[width=120mm]{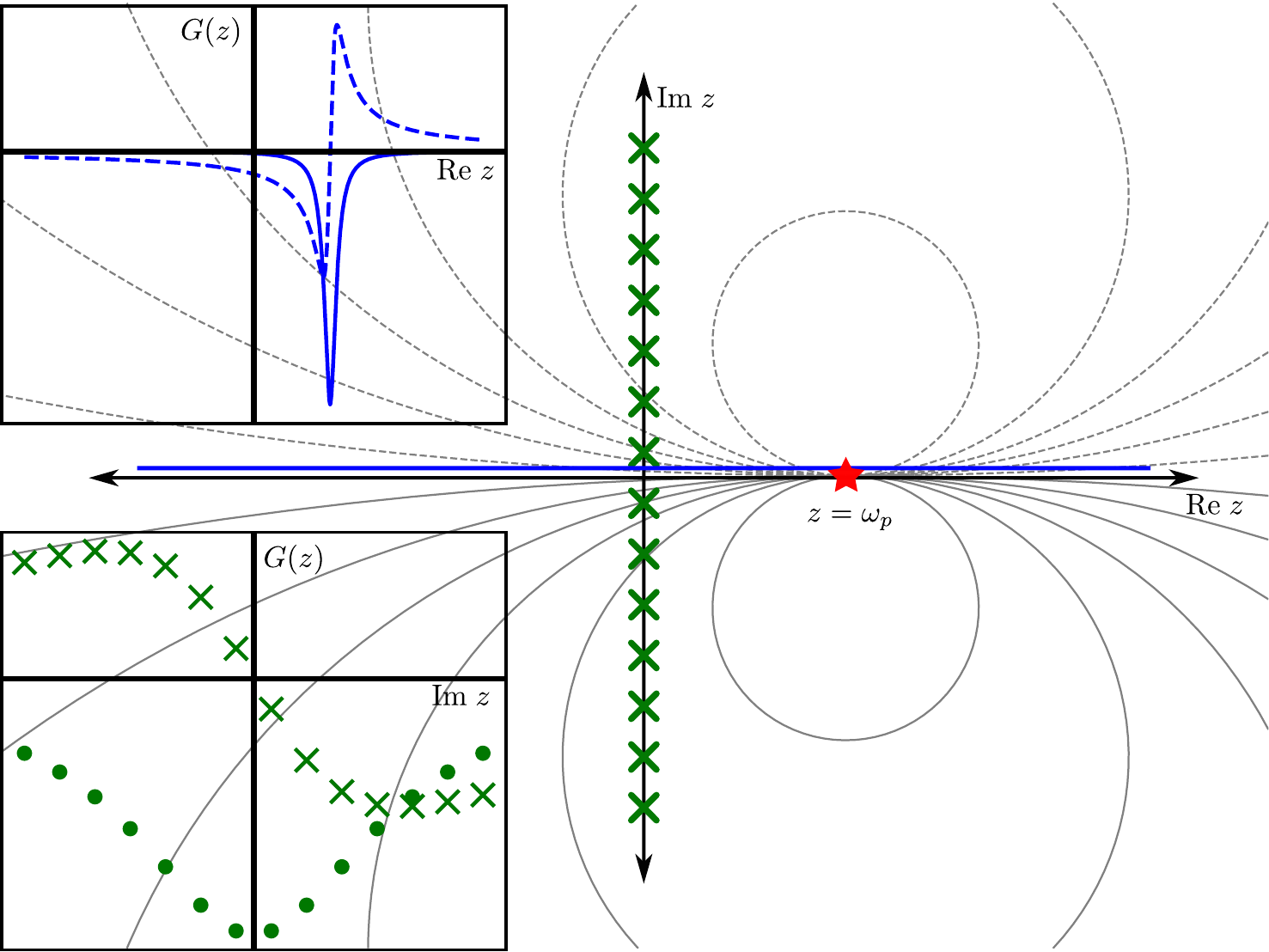}  
  \caption{\label{fig:complex-plane} Simple one-pole Green's function $G(z) = 1/(z-\omega_p)$ on the plane of complex frequencies $z$.
	The background density/contours show only $|\mathrm{Im} \, G(z)|$. 
  The pole location $\omega_p$ is marked by a red asterisk.
  The inset in the upper left shows the retarded Green's function (evaluated along the blue line on the complex plane), 
  where the full line is the imaginary part and the dashed line is the real part.
  Clearly, the imaginary part of the retarded Green's function has a peak at $\omega=\omega_p$.
  The inset in the lower left shows the Matsubara Green's function (evaluated at the green crosses on the complex plane).
  Its real and imaginary part are drawn by dots and crosses, respectively.
  }
\end{figure}

It can be shown that all physical information of the Green's function
is encoded in a \emph{spectral function} $A(\omega)$, since at
any complex frequency $z$ the Green's function is
\begin{equation}
  \label{eq:relation-gf-specdens}
  G(z) = \int_{-\infty}^\infty d\omega' \frac{A(\omega')}{z - \omega'}.
\end{equation}
By evaluating \cref{eq:relation-gf-specdens} at Matsubara frequencies $z=i\omega_n$,
we get the Matsubara Green's function,
and at $z=\omega+i0^+$ we get the retarded Green's function, as depicted in \cref{fig:complex-plane}. Generally,  \cref{eq:relation-gf-specdens} allows calculating the Green's function in the entire complex plane.
By applying the Sokhotski-Plemelj theorem (often called Weierstrass formula)
\begin{equation}
  \label{eq:weierstrass}
  \mathrm{lim}_{\epsilon\rightarrow 0^+} \int_{-\infty}^\infty \!\!dx\, \frac{f(x)}{x + i\epsilon}
  =
  -i\pi f(0)
  + \mathcal{P}\int_{-\infty}^\infty \!\!dx\, \frac{f(x)}{x}
\end{equation}
we obtain the important relation
\begin{equation}
  \label{eq:connection-spectrum-retarded-gf}
  A(\omega) = -\frac{1}{\pi} \mathrm{Im} G_R(\omega) = -\frac{1}{\pi} \mathrm{Im} G(\omega+i0^+).
\end{equation}

\section{\pade\ interpolation}\label{sec:pade}
\noindent
In \cref{sec:analytic-properties} we have seen that
the Matsubara Green's function is connected to the spectral
density through the integral relation \cref{eq:relation-gf-specdens}.
Since the spectral density contains valuable information,
it is of interest to extract the spectral
density from the Matsubara Green's function.

If the Green's function is given analytically, one may just
substitute $i\omega_n$ by $\omega+i0^+$. 
Considering that in most calculations the Green's function is
evaluated only numerically, a different approach is required.

A logical next step is to approximate the numerical values of the Green's function
by an analytical function, which can be evaluated on the real axis
through substitution as above.
This is realized in the technique of \pade\ approximants \citep{NumericalRecipes},
where a given set of data points is interpolated by a rational function.
An efficient algorithm, designed for Green's functions, was put forward 
already by \citet{VidbergSerene1977}.
Although frequently called a fit, the \pade\ approximation is
thus actually an interpolation of given data. 

\citet{VidbergSerene1977} show that using \pade\ approximants
for analytic continuation works remarkably well.
The authors however also discuss a major drawback of the method.
Namely, it is by no means clear \emph{a priori}, which
Matsubara frequencies should be selected for constructing the
interpolation. Instead they propose to compare several \pade\
approximants, where different sets of Matsubara frequencies are used.

Another problem in the construction of \pade\ approximants
arises, when the data points on the imaginary axis are subject
to stochastic uncertainty or noise. It is then impossible to still
interpolate these points by a rational function without poles in
the upper half-plane. However, noise has the potential to wreak havoc 
on the interpolation.

\begin{figure}
  \centering
  \includegraphics[width=120mm]{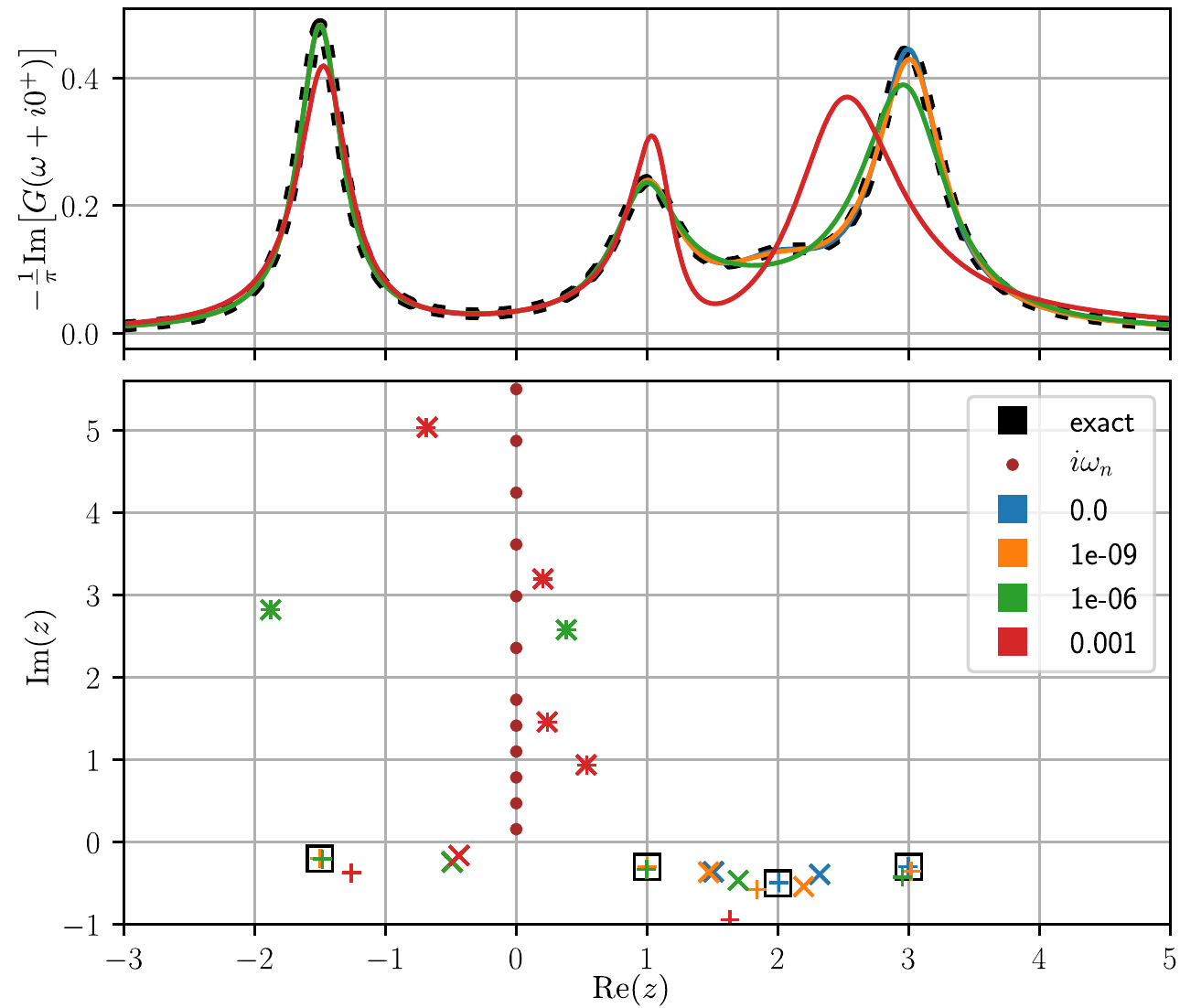}
  \caption{\label{img:pade-poles-zeros}\pade\ approximation for
  a test function with four poles (black squares). The Matsubara
  frequencies used to construct the approximant are marked by red dots.
  Reconstructed poles and zeros are marked by $+$ and $\times$ symbols,
  respectively, with the color coding the  noise level, see legend.}
\end{figure}

We illustrate this behavior 
in \cref{img:pade-poles-zeros}
by taking a simple test function with four poles. We evaluate
it on a set of Matsubara frequencies and generate four test data sets
with additional noise
by adding random numbers from a normal distribution with four different
standard deviations on different orders of magnitude. 
The \pade\ approximant constructed from the exact data (blue symbols) perfectly reproduces
the original function, its poles coincide with the true poles (black boxes)
of the analytic function. Adding noise of magnitude 10$^{-9}$ (orange symbols)
just slightly shifts some of the poles, leaving the resulting
spectral function practically unchanged. Noise of magnitude 10$^{-6}$ (green symbols)
already has a much more drastic effect. Pairs of almost canceling
poles and zeros appear on the upper half plane, which is necessary
for interpolation of the noisy test data. Further increasing the
standard deviation of the noise to 10$^{-3}$ (red symbols), a value that is
by no means uncommon in actual numerical calculations, increases
the number of pole-zero pairs and the spectral function is already 
somewhat deteriorated.

\section{Probabilistic approach - Maximum Entropy}\label{sec:maxent}
\subsection{Formulation of the analytic continuation problem}
\noindent
Commonly employed are quantum Monte Carlo simulations, where the results
intrinsically contain some uncertainty.
Although we have shown above that the \pade\ approximation can
yield good results even in the case of noisy data,
it is difficult to assess its quality in real-world cases, 
where the exact solution is unknown.
The interpolation treats data with random noise as if they were
exact. Thus, the deviation of the spectral function
from the true underlying spectral function has an element of
arbitrariness.

For this reason, most methods for analytic continuation do not try
to construct an exactly interpolating analytic function. 
Instead, one tries to answer the following question:
Assume that $N$ normal-distributed data $G_n$ have been measured 
with corresponding uncertainty $\sigma_n$.
Given a linear relation like \cref{eq:relation-gf-specdens} between the spectral function $A(\omega)$ and $G_n$,
\begin{equation}
  \label{eq:linear-relation}
  G_n = \int_{-\infty}^\infty \!\!d\omega\, K(i\omega_n, \omega)\, A(\omega) 
  \hspace{2em}
  \text{or in operator notation}
  \hspace{2em}
  G = K  A,
\end{equation}
what is the most probable spectral function $A(\omega)$ that fits the data?
One can immediately write down the
probability for measuring $G$ if the spectral function is $A(\omega)$:
\begin{equation}
  \label{eq:likelihood-function}
  \pr[G | A] = \frac{(2\pi)^{-N/2}}{\prod_i \sigma_i}
  \mathrm{exp}\big[
    -\sum_i \frac{(G_i - K  A)^2}{2\sigma_i^2}
    \big]
\end{equation}
The maximum likelihood method [cf., e.g., \citet{NumericalRecipes}] is
based on the assumption that this is numerically equivalent
to the probability of $A$ being the true spectral function, after $G$
has been measured (assuming we have no {\em a priori} information about $A$):
\begin{equation}
  \label{eq:maximum-likelihood}
  \pr [A | G] = \pr[G | A].
\end{equation}
Then, the most probable spectral function is the one that maximizes the exponent
of \cref{eq:likelihood-function}, or in other words, minimizes the merit function $\chi^2$
\begin{equation}
  \label{eq:chi2}
  \chi^2[A] = \sum_{n=1}^N \frac{[G_n - \int \!\!d\omega\, K(i\nu_n, \omega)\,A(\omega)]^2}{\sigma_n^2}.
\end{equation}
This is a general least squares problem \citep{NumericalRecipes}.
Let us for further analysis discretize the integral,
\begin{equation}
  \label{eq:discretize-integral}
  \int\!\! d\hspace{-0.5pt}\omega\, K(i\nu_n, \omega)\, A(\omega)
  \approx
  \sum_k K(i\nu_n, \omega_k) A(\omega_k) \Delta\omega_k.
\end{equation}
Only for the sake of demonstration, assume that the measurement uncertainty $\sigma_n$
is constant for all $n$ and the discretization of $\omega$ is uniform, 
$\Delta\omega_k = \Delta\omega$.
In this case the least squares problem is reduced
to the following set of linear equations:
\begin{equation}
  0 = G_n - \sum_k K_{nk} A_k \Delta\omega.
\end{equation}
Hence, the kernel determines the ``hardness'' of the problem,
and it is indicated to have a closer look at its properties.

\subsection{Properties of the kernel}
\noindent
The kernel that relates spectral functions to Green's functions in Matsubara
frequencies is
\begin{equation}
  \label{eq:kernel}
  K(i\nu_n, \omega) = \frac{1}{i\nu_n - \omega}.
\end{equation}
Here, $\omega$ is a frequency on the real axis and $i\nu_n$
are (fermionic or bosonic) Matsubara frequencies.
The properties of the kernel are best understood by
first discretizing the real-frequency axis, such that
the kernel becomes a matrix $K_{jk} = K(i\nu_j, \omega_k)$.
It is however not a quadratic matrix, since the number
of real frequencies has to be large enough to resolve all
features of the spectral function, whereas the number of Matsubara frequencies 
only has to be large enough to reach the asymptotic
region. These criteria are mutually independent and thus
it would not be justified to choose the kernel to be a quadratic matrix. Numerical necessity also often restricts the number of Matsubara frequencies.

\begin{figure}
  \centering
  \includegraphics[width=\textwidth]{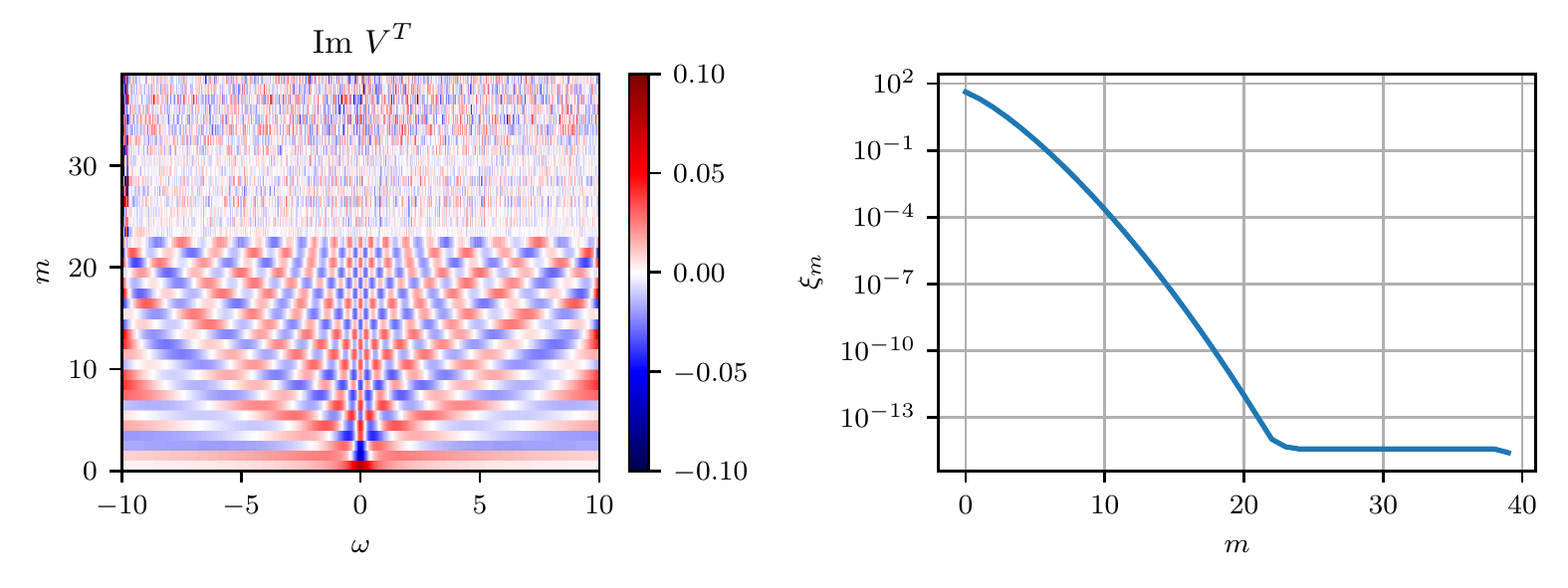}
  \caption{\label{fig:kernel-v-xi} Singular-value decomposition of the kernel
  for the analytic continuation [\cref{eq:svd}].
	Left panel: matrix of right singular vectors $V_{m}(\omega)$ ($m$: number of the  singular value).
  Right panel: log-plot of the singular values $\xi_m$.}
\end{figure}

A singular-value decomposition can be readily performed also for general
non-quadratic matrices:
\begin{equation}
  \label{eq:svd}
  K_{jk} = \sum_m U_{jm} \xi_{m} V_{km},
\end{equation}
where $U$ and $V$ are column-orthogonal matrices and $\xi$ is the
vector of singular values.
In \cref{fig:kernel-v-xi} the imaginary part of the matrix $V^T$
is shown, together with the singular values.
The columns of $V$ can be understood as basis functions for
the spectral function \citep{Shinaoka2017}.
The contribution of the $m$-th component of the spectral function
in this basis to the Matsubara data is then weighted by the $m$-th 
singular value. Since only few singular values are of significant
size, the problem of minimizing $\chi^2$ in \cref{eq:chi2} is ill-conditioned
and there is a large space of degenerate solutions.


\subsection{Regularization term or prior probability}
\noindent
If there are many completely different solutions to the problem,
this may mean that we do not have enough information to actually solve
the problem.
It may however also mean that we do not use all our knowledge
about the problem in the right way. 
Our inference has to make use of all available information,
and this can be done by means of Bayes' theorem
\begin{equation}
  \label{eq:bayes}
  \pr[A|G] = \frac{\pr[A]\;\pr[G|A]}{\pr[G]},
\end{equation}
where we want to find the spectral function $A$ (hypothesis) that maximizes the conditional
or posterior probability $\pr[A|G]$ of the hypothesis being true if
data $G$ have been measured. The likelihood function $\pr[G|A]$ has
already been defined in \cref{eq:likelihood-function}. Additionally,
with $\pr[A]$ and $\pr[G]$ we now have the prior probabilities of the hypothesis
and the data, respectively. \cref{eq:bayes} is much richer compared to
\cref{eq:maximum-likelihood} and there is hope that with Bayesian statistics
we get better results than with the maximum likelihood method.
However, we first have to model the prior probability of a spectral function, 
$\pr[A]$. This modeling is not univocal \citep{NumericalRecipes},
but a highly successful and state of the art choice of the entropic prior is
\begin{equation}
  \label{eq:entropic-prior}
  \pr[A] \propto e^{\alpha S[A]}
\end{equation}
with the entropy
\begin{equation}
  \label{eq:entropy}
  S[A] = \int \hspace{-5pt} d\hspace{-0.5pt}\omega \left[ A(\omega) - D(\omega) - A(\omega)\,\mathrm{log}\frac{A(\omega)}{D(\omega)} \right]
\end{equation}
relative to a default model $D(\omega)$. This
leads to the maximum entropy method (\maxent). 
The concept of entropy was introduced into information theory by
\citet{Shannon1948}, and \citet{Jaynes1957} established the principle of maximum
entropy as ``a method of reasoning'' in statistical mechanics.
Very successfully the maximum entropy method was employed for
deconvolution of optical data \citep{Frieden1972, Gull1978}.
After further developments and applications to image processing
\citep{Gull1984}, it was also applied to the analytical continuation
problem \citep{Silver1990, Silver1990PRL, JarrellGubernatis1996}.

Combining \cref{eq:bayes} with \cref{eq:likelihood-function}, \cref{eq:entropic-prior}, and \cref{eq:entropy},
we have to maximize the probability
\begin{equation}
  \label{eq:posterior-probability}
  \pr[A|G] = \frac{1}{\pr[G]} e^{-Q_\alpha[A]},
\end{equation}
where
\begin{equation}
  \label{eq:maxent-Q}
  Q_\alpha[A] = \frac12 \chi^2[A] - \alpha S[A]
\end{equation}
with a yet unspecified scaling hyperparameter $\alpha>0$.
The prior probability of the data, $\pr[G]$, also referred to as evidence,
is fixed, since we are working with one measured data set.
Therefore it does not depend on the spectral function and
in our considerations can be absorbed in the normalization of $\pr[A|G]$.

Instead of the least-squares problem of \cref{eq:chi2} we now face
the minimization of the functional $Q_\alpha[A]$.
The entropy, which was introduced as the prior probability of the spectral function,
now serves as a regularization term to the ill-conditioned least-squares problem.
This becomes apparent by looking at the Hessian matrix of $Q$ with respect to $A$:
\begin{equation}
  \label{eq:q-hessian}
  \frac{\partial^2 Q_\alpha}{\partial A_l\, \partial A_m} = 
  \sum_n \frac{K_{nl}\Delta\omega_l \,K_{nm}\Delta\omega_m}{\sigma_n^2}
  + \alpha \delta_{lm}\frac{\Delta\omega_l}{A_l},
\end{equation}
which is here computed by choosing a discretization of the real-frequency axis
and deriving by the value of the spectral function at these discrete points.
The first term of \cref{eq:q-hessian} contains a product $K^T K$,
thus its conditioning is even worse than that of $K$: most of its eigenvalues
are practically zero. 
However, the second
term of \cref{eq:q-hessian} adds a positive diagonal matrix, scaled by $\alpha$.
This has the effect of increasing the eigenvalues, i.\ e.\ the curvature
of $Q$ at its minimum, and thus makes it easier to actually locate the minimum.

Let us note that, instead of the entropy, one may also choose a different regularization
term. \citet{SpM} introduced the so-called sparse modeling technique,
where solutions are prioritized if they are sparse in the singular space
of the kernel. An implicit regularization effect can be achieved in
stochastic sampling methods \citep{Sandvik1998, StochReg1, StochReg2, Ghanem2020}.

\section{Technical aspects of the Maximum Entropy method}\label{sec:technical}

\subsection{Choice of the hyperparameter $\alpha$}\label{sec:choice-of-alpha}
\begin{figure}
  \centering
  \includegraphics[width=\textwidth]{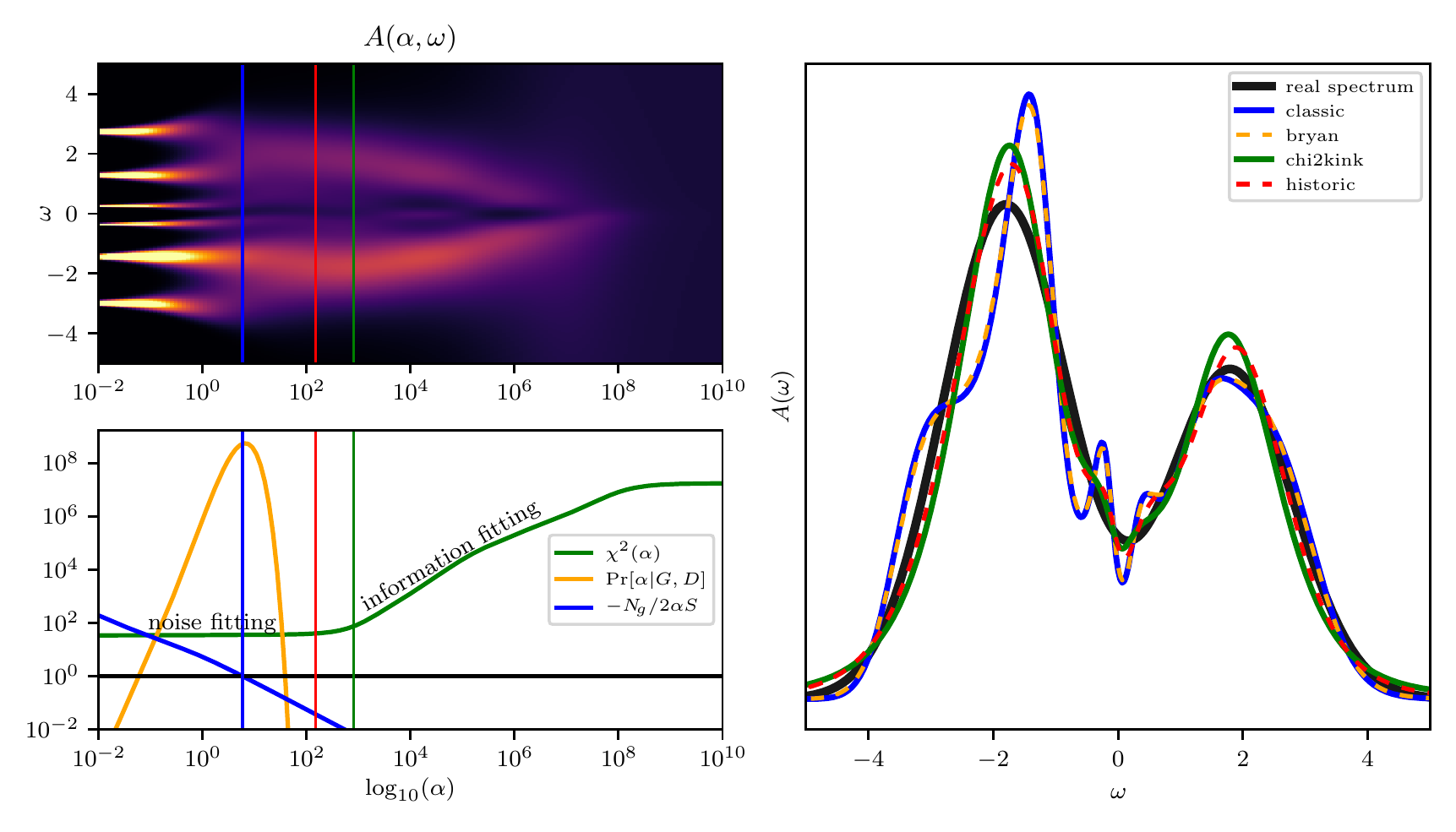}
  \caption{\label{fig:spec-alpha}Methods for determining $\alpha$.
  A model spectrum (black curve in the right panel) is transformed to Matsubara frequencies
  and normal-distributed noise with amplitude $10^{-4}$ is added. 
  Upper left panel: spectral function $A$ (color code)  obtained through
	the minimization of the \maxent\ functional $Q_\alpha$ for a range of values of $\alpha$.
	Lower left panel: illustration of the four different
  methods of determining the optimal value of $\alpha$. Here the
  red and blue vertical lines mark historic and classic \maxent, respectively (cf.\ colors in the legend of the right panel). The value of
  $\alpha$ obtained by the ``chi2kink'' method is marked by a green vertical line.
  Note that Bryan's method can not be highlighted in such a way since
  it is a weighted average over a range of $\alpha$ (in this case  the probability
  that serves as a weight is  drawn in orange.) 
	Lower right panel: Spectral functions obtained by the four different methods
	compared to the real (initial) spectrum.}
\end{figure}

\noindent
Having seen that the functional $Q_\alpha[A]$ can be minimized for $\alpha>0$,
we now have to determine how to actually choose a good value of $\alpha$.
The importance of this choice is illustrated in \cref{fig:spec-alpha}. 
We choose the asymmetric two-peak spectrum that is shown as a black
curve in the right panel. By \cref{eq:kernel} it is transformed to Matsubara
frequencies, and noise with an amplitude of 10$^{-4}$ is added.
Assuming a flat default model, 
we calculate the spectral function $A(\omega)$ that minimizes \cref{eq:maxent-Q}
for a large range of different values of $\alpha$.

The influence of
$\alpha$ is shown as a density plot in the top left panel of \cref{fig:spec-alpha}.
At very high values of $\alpha$, we recover the constant default model,
no influence of the data can be noticed. 
As $\alpha$ is decreased, the spectral weight concentrates in the middle,
then we observe the formation of two distinct peaks. 
At small values of $\alpha$, the peaks split and in the end we get six
very sharp peaks instead of two broad ones.
The number and positions of such peaks is largely determined by the noise
of the data. In our test case, we could generate the noise with a different
random seed, which would lead to a seemingly completely different spectrum
in the low-$\alpha$ limit. 

Since apparently the result of the analytic continuation is to a large extent
controlled by $\alpha$, it is extremely important to choose it in a reasonable
and automatizable way.
Various approaches have been proposed in order to tackle this problem. 

In so-called \emph{historic} \maxent, $\alpha$ was chosen in a way that the 
$\chi^2$-deviation (\cref{eq:chi2}) is approximately equal to the number
of data points on the imaginary axis \citep{Gull1989}.
It was however found to under-fit the data and both \citet{Gull1989} and \citet{Skilling1989}
argue that it is somewhat \emph{ad hoc}.

The \emph{classic} \maxent\ extends the Bayesian 
inference scheme to $\alpha$ and $D$ \citep{Skilling1989,JarrellGubernatis1996}:
\begin{equation}
  \label{eq:bayes-alpha}
  \pr[A, \alpha | G, D] \propto \pr[A | G, D, \alpha] \pr[\alpha]
  \propto \frac{e^{-Q_\alpha[A]}}{\alpha},
\end{equation}
where the scale invariant Jeffreys' prior $1/\alpha$ \citep{Jeffreys1939} 
was used for the prior probability of $\alpha$. 
Subsequently, the dependence on the spectral function in \cref{eq:bayes-alpha}
is integrated out and one obtains an explicit form for $\pr[\alpha|G, D]$.
In the bottom left panel of \cref{fig:spec-alpha} the logarithm of the posterior probability
$\pr[\alpha| G, D]$ is shown as an orange curve. Classic \maxent\ takes the spectral function
at the most probable value of $\alpha$ to be the best solution to the problem.
This value is usually obtained by setting $\partial\pr[\alpha| G, D]/\partial\alpha=0$,
which can be analytically transformed to an equation $N_g=-2\alpha S$
after a few approximations. $N_g$ is called the ``number of good measurements'', for its
definition we refer to \citet{JarrellGubernatis1996}.
In the bottom left panel of \cref{fig:spec-alpha} we plot $-N_g/2\alpha S$ as a blue line;
its intersection with 1 is the classic \maxent\ solution, marked by a blue vertical line.
The corresponding spectrum is shown in the right panel.

Closely related is \emph{Bryan's method} \citep{Bryan1990}, where one takes the average
over all values of $\alpha$, weighted by the probability:
\begin{equation}
  \label{eq:bryans-method}
  A(\omega) = \hspace{-3pt} \int \hspace{-5pt} d\hspace{-0.5pt}\alpha\, A(\alpha, \omega)\, \pr[\alpha|G, D]
\end{equation}
The results are usually very close to classic \maxent, since
the probability of $\alpha$ is sharply peaked,
see the blue and orange dashed lines in \cref{fig:spec-alpha}.
As one can clearly see, the solutions of classic and Bryan's \maxent\
may lead to unphysical extra peaks --- a drawback already noticed
by \citet{Gull1989}.

However, there is yet another approach to the decision of the best $\alpha$.
It is based on the so-called L-curve criterion \citep{Lawson1974, Lawson1995}. In the context of \maxent, it was to our knowledge first proposed and explained by \citet{Bergeron2016}
and in the following also applied by \citet{Kraberger2017}.
It is somewhat more heuristic than classic or Bryan's \maxent\ in that
it relies only on the behavior of $\chi^2$ (like historic \maxent).
For apprehending this approach, we plot $\chi^2$
as a green curve in the lower left panel of \cref{fig:spec-alpha}.
In the limit of $\alpha\rightarrow\infty$ it goes to a constant high value, 
and for $\alpha\rightarrow 0$ another constant of lower value is approached.
Following the reasoning of \citet{Bergeron2016}, there is a clear interpretation
for this behavior.
In the high-$\alpha$ limit, the (large) $\chi^2$-deviation has negligible weight,
such that the default model $D$ minimizes $Q_\alpha[A]$. What we
see is thus $\chi^2[A=D]$. On the other hand, in the low-$\alpha$ limit, 
the contribution of the entropy is reduced to enforcing positivity of the spectrum;
and $\chi^2$ slowly approaches its global minimum. The low-$\alpha$ region,
where $\chi^2$ is relatively flat, and the subsequent
region with steep increase of $\chi^2$, are coined \emph{noise-fitting} and 
\emph{information-fitting} regions \citep{Bergeron2016}.
The optimal value of $\alpha$ is situated in the transition region
between information fitting and noise fitting,
where $\mathrm{log}\chi^2(\mathrm{log}\alpha)$ has a kink. We therefore call this method to determine
$\alpha$ ``chi2kink''.
This optimal $\alpha$ is indicated by a maximum in the curvature $\chi''(\alpha)$
\citep{Bergeron2016, Kraberger2017}.
A more numerically stable and flexible approach is to fit a function
\begin{equation}
  \label{eq:fit-chi2-alpha}
  \phi(x; a,b,c,d) = a + \frac{b}{1 + e^{-d(x-c)}}
\end{equation}
to several values of $\mathrm{log}_{10}\chi^2(\mathrm{log}_{10}\alpha)$.
The curvature of $\phi$ in \cref{eq:fit-chi2-alpha} is maximal at 
$x=c-\mathrm{ln}(2+\sqrt{3})/d \approx c - 1.317/d$.
Since this usually leads to strong underfitting, 
we suggest to use a more general $x=c-f/d$ or $\alpha=10^{c-f/d}$, 
with preferred values of $f \in [2, 2.5]$.
In \cref{fig:spec-alpha} the value corresponding to $f=2$ is marked by green vertical lines.

In summary, historic \maxent\ seems to underfit only with respect
to classic \maxent, since the latter tends to grave overfitting.
\citet{Bergeron2016} note that the formula for the posterior probability
of $\alpha$ is an approximation that works well only when a
good default model is used. 

In order to get good results with classic \maxent, it has therefore
become common practice to increase the standard deviation $\sigma_n$ 
by a manually chosen factor to shift
the peak in the probability to a larger value of $\alpha$. 
Since the rescaling factor is not known \emph{a priori} 
(although a dependence on the actual noise amplitude is noticed),
this procedure introduces an additional degree of arbitrariness.

Determining the optimal $\alpha$ as the point where noise-fitting starts
and information fitting ends (``chi2kink'')
leads to considerably larger values of $\alpha$, even with respect
to historic \maxent. Despite the drawback of being less ``Bayesian'',
it does not rely on additional approximations and the noise amplitude of the
data has to be known only up to a constant prefactor.
The only remaining source of arbitrariness is the parameter $f$ mentioned
above, which controls the actual value of $\alpha$ to be accepted.
It is however restricted to a rather narrow range, whereas
the error rescaling factor in classic \maxent\ can vary by several
orders of magnitude.

At least in the present context of analytic continuation, we are therefore
convinced that the best method of determining $\alpha$ is extracting it
as the border between noise- and information-fitting.

\subsection{Elimination of unphysical features: Preblur}\label{sec:preblur}
\noindent
Even though we are now able to find the ``best'' value of $\alpha$,
the spectrum corresponding to it may show some undesirable artifacts.
In the right panel of \cref{fig:spec-alpha} we can see
that all solutions have an enlarged curvature around $\omega=0$ and
side peaks emerge upon lowering $\alpha$.
An analogous tendency was observed already in the original application
of \maxent\ in image processing, where it manifested itself in
huge peaks in the intensity distribution. In order to address this problem,
 \citet{Skilling1991} showed that the entropy does not need to be evaluated from
the spectral function directly, but from a ``hidden'' function $h(\omega)$
that is related to the spectral function in a linear way
\begin{equation}
  \label{eq:hidden-matrix}
  A = \Omega h.
\end{equation}
In standard \maxent, $\Omega$ is just unity.
Although initially requiring $\Omega$ to be an orthogonal matrix,
\citet{Skilling1991} subsequently assumed a Gaussian convolution
$A=g_b\!\ast\!h$
with great success and named the technique ``preblur''.
In the latter case, $\Omega$ is a matrix that contains a Gaussian
function $g_b(\omega)=\mathrm{exp}[-\omega^2/2b^2] / \sqrt{2\pi}b$ in each row,
i.e.\ $\Omega_{ij} = g_b(\omega_i \!-\! \omega'_j)$.
In the context of analytic continuation,
preblur was introduced by \citet{Kraberger2017},
although already \citet{Sandvik1998} uses a somewhat similar smoothing algorithm.

The functional of \cref{eq:maxent-Q}, which we have to minimize, thus becomes
\begin{equation}
  \label{eq:preblur-functional}
  Q_{\alpha b}[h] = \frac{1}{2}\chi^2[g_b\! \ast\! h] - \alpha S[h].
\end{equation}
There we now have, after $\alpha$, a second hyperparameter $b$
that controls the width of the Gaussian $g$.

Let us now have a closer look at the $\chi^2$-term of \cref{eq:preblur-functional}.
It sums up the difference of the fit ``$KA$'' to the data $G$.
Explicitly the fit function is now
\begin{align}
  \label{eq:preblur-fit}
  K A &\equiv \int_{-\infty}^\infty \hspace{-5pt} d\omega\, K(i\nu_n, \omega)\, A(\omega) \notag\\
  &= \int_{-\infty}^\infty \hspace{-5pt} d\omega\, K(i\nu_n, \omega)\, \int_{-\infty}^{\infty}
    \hspace{-5pt} d\omega'\, g_b(\omega-\omega')\, h(\omega')\notag\\
    &= \int_{-\infty}^\infty \hspace{-5pt} d\omega \Big[ 
        \int_{-\infty}^\infty \hspace{-5pt} d\omega'\, K(i\nu_n, \omega')\, g_b(\omega - \omega')
    \Big] h(\omega)\notag\\
  &\equiv \tilde{K} h.
\end{align}
This means that the hidden spectral function is related to the Matsubara
Green's function in a similar way as the spectral function, the only
difference being a ``blurred'' kernel $\tilde{K}$. 
Therefore, preblur is realized by minimizing the functional $Q$
with a slightly modified kernel, which then yields the hidden spectral
function $h$. The spectral function $A$ itself is subsequently obtained by
the convolution of $h$ with the same Gaussian $g_b(\omega)$ as above. 

\begin{figure}
  \centering
  \includegraphics[width=\textwidth]{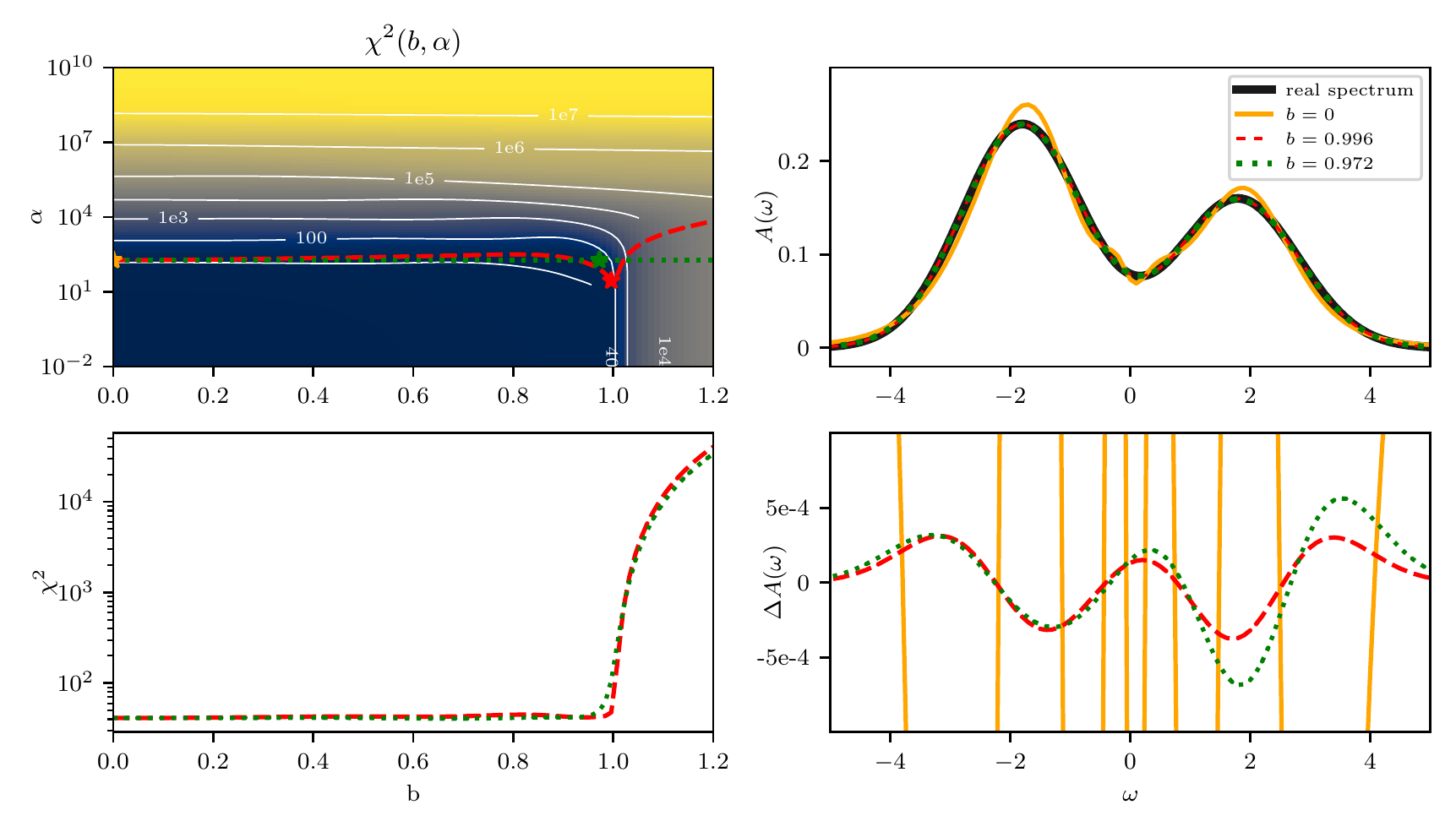}
  \caption{\label{fig:preblur} Analyzing the preblur-parameter $b$. Upper left panel: $\chi^2$-deviation (color code and isolines) of
  the MaxEnt fit as a function of the hyperparameters $b$ and $\alpha$. The red dashed line marks the optimal $\alpha$ at each $b$, the
  green dotted line shows the optimal $\alpha$ at $b=0$ for comparison. Lower left panel: $\chi^2$ along the
  red dashed and green dotted lines in the upper left panel. Upper right panel: The true spectral function (black),
  and its \maxent\ reconstructions at the points marked by orange, red and green asterisks in the upper left panel, 
  drawn in the respective colors. The lower right panel shows the differences of these reconstructions
  to the true spectrum. The orange $\Delta A(\omega)$ (corresponding to $b=0$) is as large as $10^{-2}$.
  }
\end{figure}

Although the minimization of $Q$ itself does not become more difficult
by introducing preblur, we now face the problem that the blur width $b$
is not known \emph{a priori}, similar as the entropy-scaling parameter $\alpha$.
However, knowing that we can determine a good value for $\alpha$ by
analyzing $\chi^2(\alpha)$, it is now worth to analyze the $\chi^2$-deviation
as a function of both $\alpha$ and $b$. This is done in the upper left panel
of \cref{fig:preblur}. Over a large range of $b$, $\chi^2$ depends only on $\alpha$.
Only after a sharp border (at $b\approx 1$ in our example)
the $\chi^2$-deviation increases steeply with $b$,
which can be seen best in the lower left panel of \cref{fig:preblur}.
In other words, increasing $b$ up to a certain value 
does not change the quality of the fit (but potentially the result.) 
Only afterwards it is suddenly impossible to get a good fit.
A heuristic explanation of this is that convolving the kernel with a Gaussian
permits only peaks with a minimal width of $2b$ in the spectrum.
Once this exceeds the ``real'' width of one of the peaks in the spectrum,
a fit becomes impossible. This observation is corroborated by the fact
that we find the maximal permissive $b$ to be close to $1$ in \cref{fig:preblur},
and the actual spectrum (black curve) consists of two Gaussians with standard deviation 1.

The observations made in the upper left panel of \cref{fig:preblur} can be used to propose
the following algorithm for the determination of the hyperparameters:
Determine the optimal $\alpha$ for different values of $b$, starting from 0 and increasing. 
Accept the last value of $b$ where $\chi^2$ at the optimal $\alpha$ is smaller than,
e.\ g., 1.5 times $\chi^2$ at the optimal $\alpha$ for $b=0$.
The optimal values of $\alpha$ are marked as a red dashed line in the upper left panel of \cref{fig:preblur},
and the thus determined $(\alpha, b)$ is marked by a red asterisk.
Since the optimal values of $\alpha$ hardly change in the 
reasonable range for $b$, one may also use a simplified algorithm:
Determine the optimal $\alpha$ for $b=0$, then increase $b$, keeping always the same $\alpha$.
If $\chi^2$ exceeds, e.\ g., 1.5 times its value at $b=0$, accept the last value of $b$
before this as the best value. The corresponding $(\alpha, b)$ is marked in \cref{fig:preblur}
by a green asterisk.

As we can see in the upper right panel, the corresponding red and green curves
obtained by analytic continuation with these two $(\alpha, b)$
are practically indistinguishable. In the lower right panel, the differences of these reconstructed
spectra are plotted and show that the deviation from the actual result is now satisfactorily small.

Let us note in passing that the above described unphysical wiggles around $\omega=0$ are most prominent in
spectral functions similar to our example case. In cases where the spectral function
shows a gap or a peak at $\omega=0$, preblur is usually not required. 

\subsection{\maxent\ for offdiagonal elements}
\noindent
The considerations above were done for positive definite spectral functions
with a finite norm, typically one: $\int d\omega A(\omega) = 1 > 0$. 
This property is fulfilled by diagonal elements of correlation functions.
In some cases it is however necessary to do an analytic continuation of
offdiagonal elements, which have 
zero norm $\int d\omega A(\omega) = 0$, as a direct consequence of the fermionic
anticommutation relation. A vanishing norm is not consistent with a positive
definite function, thus the spectral function cannot have a definite sign for offdiagonal elements.
Clearly the definition of the entropy, \cref{eq:entropy}, has to be adapted
to this new circumstance \citep{Kraberger2017}.
At this point, let us mention that the kernel $K(i\omega_n, \omega)$
is the same for diagonal and offdiagonal elements.

Let us, for offdiagonal components, write the spectrum as
\begin{equation}
  \label{eq:A-offdiag}
  A(\omega) = A^+(\omega) - A^-(\omega),
\end{equation}
where both $A^+$ and $A^-$ are positive definite and have the same norm. 
Then we can write the entropy as\footnote{Usually, two default models $D^+$ and $D^-$
are introduced, and assumed to be equal at a later point. For the sake of brevity,
we omit this step.}
\begin{align}
  \label{eq:entropy-posneg}
  S[A^+, A^-] = \int\hspace{-5pt} d\hspace{-0.5pt}\omega \Big[
    &A^+(\omega) - D(\omega) - A^+(\omega)\, \mathrm{log} \frac{A^+(\omega)}{D(\omega)}\notag\\
      + &A^-(\omega) - D(\omega) - A^-(\omega)\, \mathrm{log} \frac{A^-(\omega)}{D(\omega)}
    \Big]
\end{align}
Instead of treating $A^+$, $A^-$ and $A$ as completely independent, we can use
\cref{eq:A-offdiag} to eliminate $A^-$ and write
\begin{equation}
  Q_\alpha[A, A^+] = \frac12 \chi^2[A] - \alpha S[A, A^+].
\end{equation}
Since we are searching for a minimum of $Q$ with respect to both $A$ and $A^+$, 
we can use
\begin{equation}
  \frac{\partial Q_\alpha[A, A^+]}{\partial A^+} 
  = -\alpha \frac{\partial S[A, A^+]}{\partial A^+} = 0
\end{equation}
to eliminate also $A^+$. Substituting   $A^-=A- A^+$ in  \cref{eq:entropy-posneg} and taking the derivative,  $A^+$ and $A^-$ can be expressed by $A$ as
\begin{equation}
  A^\pm[A] = \frac{\sqrt{A^2 + 4D^2} \pm A}{2}.
\end{equation}
Inserting this back in \cref{eq:entropy-posneg}, we arrive at the so-called
\emph{positive-negative entropy}
\begin{equation}
  \label{eq:entropy-posneg-2}
  S[A] = \hspace{-3pt} \int \hspace{-5pt} d\hspace{-0.5pt}\omega\, A(\omega)\, \mathrm{log} \frac{A^+[A](\omega)}{D(\omega)}.
\end{equation}
As a reasonable choice for the default model, \citet{Kraberger2017} propose to include the
diagonal elements of the spectrum in the following way:
\begin{equation}
  \label{eq:model-offdiag}
  D_{ll'} = \sqrt{A_{ll}A_{l'l'}} + \varepsilon.
\end{equation}
In our implementation, it is not necessary to add a small number $\varepsilon$.

Also methods that continue matrix-valued Green's functions as a whole, 
and not component-wise, have been developed \citep{Kraberger2017, Sim2018}.

\section{Python package for analytic continuation: \texttt{ana\_cont}}\label{sec:package}
\noindent
In the previous sections we have presented two methods for analytic continuation,
the \pade\ approximation and the maximum entropy method.
Although both of them are well-established and have been implemented before
[e.g.~by \citet{Bergeron2016}, \citet{Levy2017}, \citet{Kraberger2017}],
we here introduce  yet another code for analytic continuation.
In contrast to other implementations, it is a Python package.
The advantage of this approach is increased maintainability. 
There are too many ways of doing analytic continuation to encode all
of them into a few command line arguments.
The \texttt{ana\_cont} package is available open source at
\anaconturl.

\subsection{Package structure}
\noindent
The central class of \verb=ana_cont= is \verb=AnalyticContinuationProblem=,
which holds problem-specific information: real-frequency grid, imaginary frequency or time grid,
data on imaginary axis, inverse temperature, and type of the kernel,
see \cref{tab:args-probl}. It has a method \verb=solve=,
which takes several keyword arguments. The most important one is \verb=method=. Currently, 
only two methods, \verb=pade= and \verb=maxent_svd= are implemented. 
Calling \verb=solve= creates an instance of a solver object. In case of
\verb|method='maxent_svd'|, the class \verb$MaxentSolverSVD$ is instantiated. 
The class contains all necessary functions to solve the analytic continuation problem defined before.
Possible keyword arguments are listed in \cref{tab:args-solve}.

\begin{table}
  \centering
  \begin{tabularx}{0.95\textwidth}{l X}
    keyword & value\\
    \hline
    \texttt{im\_axis} & numpy array, shape $(N_{\text{mats}},)$\\
    \texttt{re\_axis} & numpy array, shape $(N_{\text{real}},)$\\
    \texttt{im\_data} & numpy array, shape $(N_{\text{mats}},)$, float or complex\\
    \texttt{kernel\_mode} & one of \texttt{'freq\_fermionic'}, \texttt{'freq\_bosonic'}, \texttt{'time\_fermionic'}, \texttt{'time\_bosonic'}\\
    \texttt{beta} & float (only required for ``time'' kernels)
  \end{tabularx}
  \caption{\label{tab:args-probl} Keyword arguments for initializing the \texttt{AnalyticContinuationProblem}.}
\end{table}

\begin{table}
  \centering
  \begin{tabularx}{0.95\textwidth}{l X}
    keyword & value\\
    \hline
    \texttt{method} & \texttt{'maxent\_svd'}\\
    \texttt{optimizer} & one of \texttt{'scipy\_lm'}, \texttt{'newton'}\\
    \texttt{alpha\_determination} & one of \texttt{'historic'}, \texttt{'classic'}, \texttt{'bryan'}, \texttt{'chi2kink'}\\
    \texttt{model} & default model: numpy array, shape $(N_{\text{real}},)$\\
    \texttt{stdev} & error bars: numpy array, shape $(N_{\text{mats}},)$\\
    \texttt{covar} & covariance matrix: numpy array, shape $(N_{\text{mats}},N_{\text{mats}},)$\\
    \texttt{alpha\_start} & float, default $10^9$\\
    \texttt{alpha\_end} & float, default $10^{-3}$\\
    \texttt{alpha\_div} & float, default $10$\\
    \texttt{fit\_position} & float, default 2.5\\
    \texttt{interactive} & boolean, default \texttt{False}\\
    \texttt{preblur} & boolean, default \texttt{False}\\
    \texttt{blur\_width} & float
  \end{tabularx}
  \caption{\label{tab:args-solve} Possible keyword arguments for \texttt{AnalyticContinuationProblem.solve()} when using \maxent.}
\end{table}

The kernel for the continuation problem at hand is provided by the class \verb|Kernel|.
It handles real-frequency discretization of the kernel, preblur,
and rotation to the eigenbasis of the covariance matrix if provided.

Furthermore there is a class \verb|GreensFunction|, whose main purpose is to construct
a full complex-valued Green's function out of a given spectrum, 
by applying the Kramers-Kronig relation \cref{eq:relation-gf-specdens}.

We want to stress that this package does not contain a ``main program'' that can access all features
by just specifying some parameters. While this may sound rather discouraging, 
we believe that in fact this \emph{reduces} the required amount of work both for
the user and for the maintainer. On the user-side, a working script can be composed of as few as 10 lines
of Python code; example scripts can be found in the GitHub repository.
There is full freedom in which way to provide the imaginary-axis data: 
They can be read from a favorite file format, or calculated on-the-fly. 
Computed real-axis data can be plotted or further processed without the detour of file-IO. 
Thus the present analytic continuation code can be easily embedded in 
complex postprocessing procedures. 
Also for the developers this principle is a large gain, since they do not have to
take care of a sophisticated user interface and flow control.
Since knowledge and experience of analytic continuation are absolutely
necessary to use any analytic continuation program, we do not see a
principal downside to this approach. 

However, for the casual user,  we have cast
in scripts a few standard procedures for analytic continuation with a graphical user interface (GUI). Thus, most of the standard analytic
continuations of bosonic and fermionic Green's functions by \maxent\ and \pade\ 
can be done even without knowledge of Python.
Tutorials for the GUI can be found in the wiki of our repository on GitHub: {
\wikiurl}

\subsection{Numerics}
\noindent
Analytic continuation is, compared to large quantum Monte Carlo simulations or Bethe-Salpeter equation inversions,
a numerically rather inexpensive task. Therefore the performance of the code
does not have to be the main goal. The \verb=ana_cont= package
is written entirely in Python, partly trading performance for flexibility and readability,
while still keeping the possibility of later optimization.
Nevertheless several actions have been taken in order to make the code reasonably fast --
a typical \maxent\ analytic continuation takes just O(1) seconds on an average desktop computer. 

\subsubsection{Frequency-space discretization.}
\noindent
For a numerical treatment it is necessary to discretize the real-frequency axis:
$\omega \rightarrow \omega_i$. For an arbitrary function $F(\omega)$ we then define
$F_i \equiv F(\omega_i)$. Integrals over frequency space are converted to sums by
$\int d\omega F(\omega) \approx \sum_i F_i \Delta_i$, where $\Delta_i$ is the width of 
the frequency interval centered at $\omega_i$. 

\subsubsection{Singular value decomposition.}
\noindent
One of the most important steps of \maxent\ is the minimization of the functional
$Q_\alpha[A]$ in the space of spectral functions. If the real-frequency axis
is discretized into, say, 1000 intervals, the space of solutions has 1000 dimensions.
This is too large for a deterministic solver, and a Monte Carlo method
is needed \citep{Sandvik1998}.
However, \citet{JarrellGubernatis1996} perform a singular value decomposition of the kernel
and thereby achieve a large reduction of dimensions, e.\ g.\ from 1000 to 20 in typical cases.
The optimization problem can then be solved by deterministic methods
like Newton root finding or the Levenberg-Marquart algorithm.
The \verb=MaxentSolverSVD= follows the strategy of singular value decomposition
of the kernel, as defined in \cref{eq:svd}.
The vector $\xi$ of singular values is truncated
such that only values larger than a certain threshold ($10^{-10}$ in our case) are kept.
In the singular space the spectrum is parameterized through
\begin{equation}
  \label{eq:singular-space-param}
  A_j = D_j \mathrm{exp} \sum_m V_{jm} u_m
\end{equation}
as proposed by \citet{JarrellGubernatis1996}.

\subsubsection{Minimization problem.}
\noindent
The above definitions are inserted into \cref{eq:maxent-Q}, and after a
few intermediate steps the stationary condition $\partial Q_\alpha/\partial A_m=0$ leads to
\begin{equation}
  \label{eq:root-fun}
  f_m(u) \equiv \alpha u_m + \xi_m \sum_k E_k U_{km} 
  \left( \sum_l K_{kl} A_l(u) \Delta_l - G_k \right) = 0, 
\end{equation}
which has to be solved for $u_m$. 
Root finding algorithms work better, when the derivative of the root
function is known analytically. Therefore, we take the derivative
of $f$ with respect to $u$ to get the Jacobian $J$:
\begin{equation}
  \label{eq:jacobian}
  \frac{\partial f_m}{\partial u_i} \equiv J_{mi} = \alpha \delta_{mi}
  + \xi_m \sum_k U_{km} E_k \sum_l K_{kl} \Delta_l A_l(u) V_{li}
\end{equation}
The expressions \cref{eq:root-fun} and \cref{eq:jacobian} can be put in
a form that is more efficient for numerical evaluation by making the following definitions:
\begin{subequations}
  \begin{eqnarray}
    w_l &\equiv& \mathrm{exp} \left( \sum_m V_{lm} u_m \right) \\
    W_{ml} &\equiv& \sum_{km} E_k U_{km} \xi_m U_{kn} \xi_n V_{ln} \Delta_l D_l \\
    W_{mil} &\equiv& W_{ml} V_{li}\\
    B_m &\equiv& \sum_k U_{km} \xi_m E_k G_k
  \end{eqnarray}
\end{subequations}
Then we have
\begin{subequations}
  \begin{eqnarray}
    f_m &=& \alpha u_m + \sum_l W_{ml} w_l - B_m \\
    J_{mi} &=& \alpha \delta_{mi} + \sum_l W_{mil} w_l,
  \end{eqnarray}
\end{subequations}
where the quantities $W_{ml}$, $W_{mil}$, and $B_m$ are precomputed
in order to minimize the number of matrix multiplications during the optimization
procedure. Please note that we do \emph{not} use the Einstein summation convention.
For the optimization, 
the Levenberg-Marquart implementation of \verb=scipy.optimize.root=
can be used by passing \verb|optimizer='scipy_lm'| to the solver.
However, we find simple Newton root finding to be much
faster and numerically stable. This is activated by setting
\verb|optimizer='newton'|.

\subsubsection{Offdiagonal elements.}
\noindent
While the above formulas are used for diagonal elements of correlation functions, 
with only very small changes we arrive at a different form that can be used
for offdiagonal elements.
The singular-space parameterization becomes \citep{Kraberger2017}
\begin{equation}
  A_j = D_j  \left[ \mathrm{exp}\left(\sum_m V_{jm}u_m\right)
                  - \mathrm{exp}\left(-\sum_m V_{jm}u_m\right) \right]
      = D_j \left( w_j - 1/w_j \right)
\end{equation}
and the minimization problem is set by
\begin{subequations}
  \begin{eqnarray}
    f_m &=& \alpha u_m + \sum_l W_{ml} (w_l - 1/w_l) - B_m \\
    J_{mi} &=& \alpha \delta_{mi} + \sum_l W_{mil} (w_l + 1/w_l).
  \end{eqnarray}
\end{subequations}

%
\subsubsection{Determination of optimal $\alpha$.}
\noindent
In \cref{sec:choice-of-alpha} we have discussed various ways of
determining a good value of the hyperparameter $\alpha$. 
All of these methods are implemented in \verb|ana_cont| and can be
used by setting the solver keyword \verb|alpha_determination| to
one of \verb|'historic'|, \verb|'classic'|, \verb|'bryan'| or,
as recommended, \verb|'chi2kink'|.

From the algorithmic point of view, all these methods have in common
that the optimization is done for several values of $\alpha$, where
one starts at a very high value of, e.g., $\alpha=10^{12}$. 
If \verb|alpha_determination='chi2kink'|,
this can be adapted by setting the keyword \verb|alpha_start|. 
One should use a value that leads to a solution very close to the default model.
Subsequently $\alpha$ is decreased by a constant factor \verb|alpha_div| in each step,
which has the advantage that the solution of the previous step can be used
as a starting point.
After reaching the smallest value of $\alpha$ (\verb|alpha_end|) the function of
\cref{eq:fit-chi2-alpha} is fitted and the parameter $f$ is specified by
the keyword \verb|fit_position|.
One has to be 
especially careful about the choice of $\alpha$ when continuing
offdiagonal elements of correlation functions. With a spectrum that is not
positive definite, one has more degrees of freedom for noise fitting.
Therefore in the double logarithmic plot of $\chi^2(\alpha)$ 
the noise fitting region may not any more be a plateau, but rather another,
less steep linear slope. The optimal $\alpha$ has to be chosen at the point, where
the slope changes.

%

\subsubsection{Preblur}
\noindent
Maxent with preblur, as described in \cref{sec:preblur} is implemented in \texttt{ana\_cont}.
Preblur is activated by passing \verb|preblur=True| to the solver,
and specifying the parameter $b$ by \verb|blur_width=|$b$. 
Our recommended workflow is to always first do an analytic continuation
with $b\!=\!0$, and then increase it step-by-step.
The final value $b^\ast$ should be large enough
to eliminate spurious features in the spectrum, but still small enough
so that the limit $\lim_{\alpha\rightarrow 0} \mathrm{log} \chi^2(\alpha)$
is not significantly increased with respect to $b=0$.

\subsection{\maxent\ for susceptibilities}
\noindent
The analytic continuation of susceptibilities, i.\ e.\ bosonic Green's functions, 
from the imaginary to the real axis is a frequently required task.
In \verb|ana_cont| it can be achieved by initializing the \verb|AnalyticContinuationProblem|
with \verb|kernel_mode='freq_bosonic'|. 
Like in the case of fermionic correlation functions,
a relation that connects the values of the susceptibility on the real and imaginary axis is required. 
A susceptibility $\chi(\omega)$ is a response function, therefore causality implies that its 
imaginary part is anti-symmetric in frequencies, 
$\chi(\omega) = \chi^\ast(-\omega)$. 
Inserting this fact in the Kramers-Kronig relation
\begin{equation*}
  \mathrm{Re}\left[\chi(i\omega_n)\right] = 
  \frac{1}{\pi} \int_{-\infty}^\infty \hspace{-5pt}d\hspace{-0.5pt}\omega\, \frac{\mathrm{Im}\!\left[ \chi(\omega) \right]}{\omega - i\omega_n}
\end{equation*}
yields
\begin{equation*}
  \mathrm{Re}\left[\chi(i\omega_n)\right] = 
  \frac{2}{\pi} \int_0^\infty \hspace{-5pt} d\hspace{-0.5pt}\omega\, \frac{\omega}{\omega^2 + \omega_n^2} \,
    \mathrm{Im}\! \left[ \chi(\omega) \right], 
\end{equation*}
connecting the susceptibility at real frequencies with its values at the bosonic Matsubara frequencies $\omega_n$.\\ 
Since the factor $\omega/(\omega^2 + \omega_n^2)$ is divergent for $\omega=\omega_n=0$, it is
common to introduce another function 
\begin{equation}
  S(\omega) \equiv \frac{2}{\pi} \frac{\mathrm{Im} \chi(\omega)}{\omega},
\end{equation}
such that we can now write
\begin{equation}
  \label{eq:relation-bosonic}
  \mathrm{Re}\!\left[\chi(i \omega_n)\right] = \int_0^\infty\hspace{-5pt} d\hspace{-0.5pt}\omega\, K_b(\omega_n, \omega)\, S(\omega)
\end{equation}
with a \emph{kernel} 
\begin{equation}
  \label{eq:kernel-bosonic}
  K_b(\omega_n, \omega) = \frac{\omega^2}{\omega_n^2 + \omega^2}
\end{equation}
that is well-defined for all values of $\omega$ and $\omega_n$. 
Correspondingly the result of the analytic continuation that is
returned by the solver of \verb|ana_cont| is $S(\omega)$.

Our \verb|ana_cont|  code has been used extensively for the analytic continuation
of susceptibilities by \citet{Geffroy2019}.

\subsection{\maxent\ for self-energies}
\noindent
Besides one-particle Green's functions and susceptibilities, 
it often is necessary to analytically continue self-energies from Matsubara to real frequencies.
Self-energies essentially have the same analytic properties as one-particle Green's functions \citep{Luttinger61}.
The difference is that the high-frequency limit is a non-zero constant $\Sigma_0$ (the so-called Hartree term), 
and the imaginary part decays
like $\Sigma_1/(i\omega_n)$ as opposed to $1/(i\omega_n)$ in case of the Green's function. 
$\Sigma_0$ and $\Sigma_1$ are called the zeroth and first moments of the self-energy
and can be calculated from density matrices \citep{Wang2011}. The zeroth moment
is frequently referred to as Hartree term.
Note that $\omega_n$ are fermionic Matsubara frequencies here. 
In absence of particle-hole symmetry, the spectral representation is
\begin{equation}
  \Sigma(i\omega_n) = \Sigma_0 + \Sigma_1 \int_{-\infty}^{\infty} \hspace{-5pt} 
    d\hspace{-0.5pt}\omega\,\frac{a(\omega)}{i\omega_n - \omega},
\end{equation}
where we defined the spectrum $a$ of the self-energy as
\begin{equation}
  a(\omega) \equiv -\frac{1}{\pi} \mathrm{Im} \frac{\Sigma(\omega) - \Sigma_0}{\Sigma_1}.
\end{equation}
Thus, we can write the analytic continuation problem for the self-energy in a form very similar to \cref{eq:relation-bosonic}:
\begin{equation}
  \label{eq:relation-selfenergy}
  \frac{\Sigma(i\omega_n)-\Sigma_0}{\Sigma_1} = \int_{-\infty}^{\infty}\hspace{-0.5pt} 
    d\hspace{-0.5pt}\omega\, K_f(\omega_n, \omega)\, a(\omega)
\end{equation}
with the fermionic kernel
\begin{equation}
  \label{eq:kernel-fermionic}
  K_f(\omega_n, \omega) = \frac{1}{i \omega_n - \omega}.
\end{equation}
We do not see any necessity for normalizing the self-energy by division
through its first moment, but merely keep this notation for consistency
with the literature.
The solver of \verb|ana_cont| then returns $a(\omega)$.
From this we can construct the imaginary part of the retarded self-energy
by inverting \cref{eq:relation-selfenergy}:
\begin{equation}
  \mathrm{Im} \Sigma_R(\omega) = -\pi \Sigma_1 a(\omega).
\end{equation}
If one wants to use the retarded self-energy for the calculation of
the retarded Green's function, it is necessary to also obtain the real part
of the retarded self-energy,
which is given by the Kramers-Kronig relation \cref{eq:relation-gf-specdens}.
This is implemented in the \verb|kkt| method of the \verb|GreensFunction| class.

\begin{figure}
  \centering
  \includegraphics[width=\textwidth]{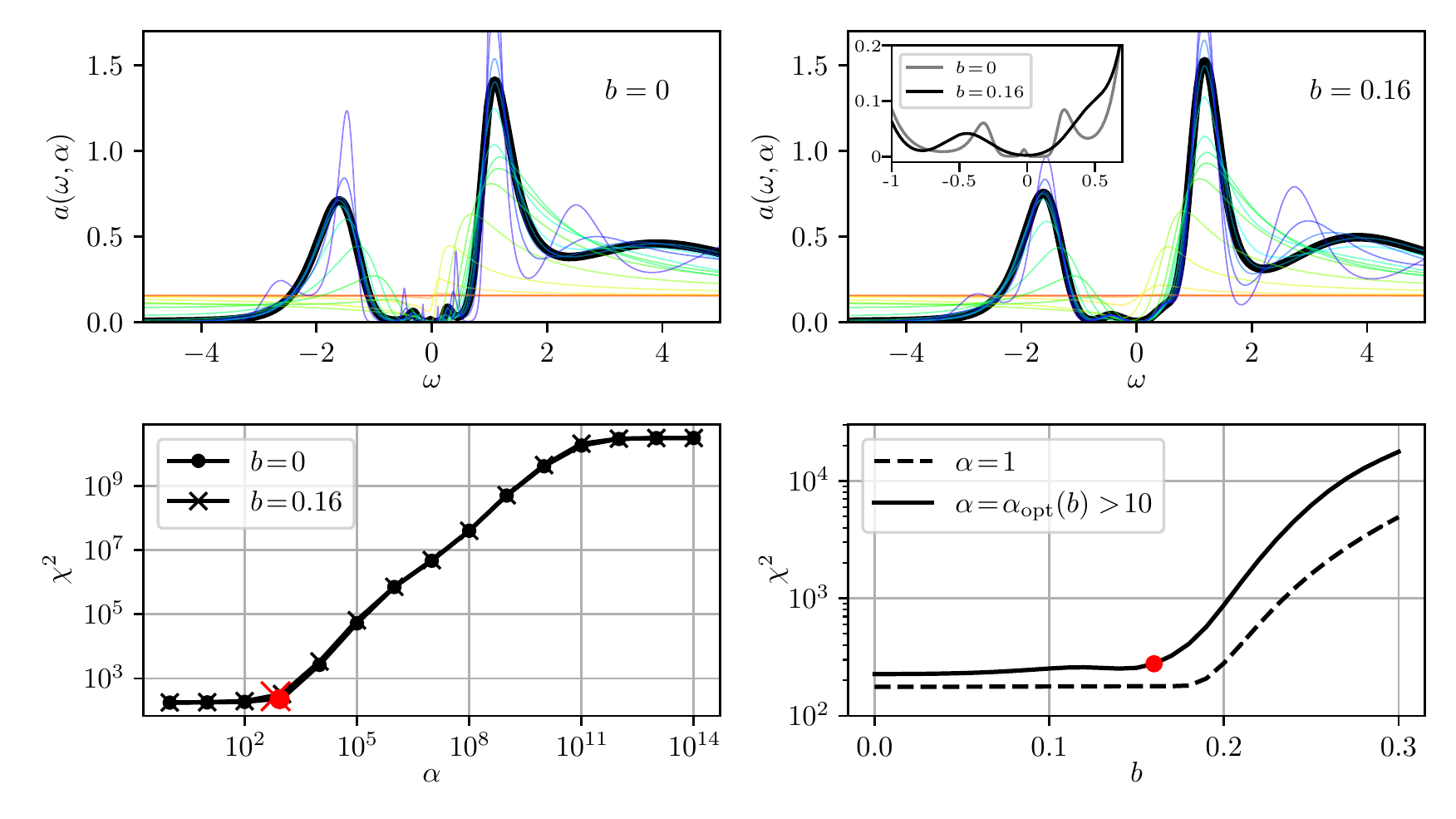}
  \vspace{-1cm}
  \caption{\label{fig:preblur-svo}
	Analytic continuation of the DMFT self-energy of the V-$t_{2g}$ orbitals of SrVO$_3$ \citep{Si2020X},
    which was obtained with symmetric improved estimators
    for extra precision \citep{Kaufmann2019}.
    \emph{Upper row}: Spectrum of the imaginary part of the self-energy spectrum $a(\omega)=- {\rm Im}\Sigma (\omega)/\pi$
    for SrVO$_3$ without [left] and with [right] preblur. The figures show optimization results
    for several different values of $\alpha$.
    Red corresponds to $\alpha=10^{13}$, and as the color changes into blue,
    the value is lowered by a factor of $10$ in every step.
    The final result, at optimal $\alpha$, is drawn in black.
    Clearly, for the highest values of $\alpha$, we recover the constant default model.
    \emph{Lower row}: Behavior of $\chi^2$
    as a function of $\alpha$ [left] and $b$ [right]
    for the \maxent\ analytic continuation of SrVO$_3$.
    The values taken as optimal are highlighted in red.}
\end{figure}

For a real-world case (SrVO$_3$) the procedure of analytic continuation
of a self-energy, using \maxent\ and preblur, is illustrated
in \cref{fig:preblur-svo}.

\section{Installation and Usage}\label{sec:installation}
\subsection{Python 3}
\noindent
As a prerequisite for using the \verb|ana_cont| package,
you need to have Python 3 installed on your computer.
This is possible via the official package repositories of most
Linux distributions. After installing Python 3, it may also be
necessary to install the Python package manager pip (again, from the official package
repositories of your linux distribution).
If necessary, you can use pip to install missing Python packages\footnote{i.e.,
numpy, scipy, matplotlib, h5py, PyQt5, Cython}:
\begin{verbatim}
pip install --user <package_name>
\end{verbatim}
Make sure that your Python 3 environment is actually activated
and pip also refers to Python 3. 
In the Anaconda\footnote{Can be downloaded from \url{https://www.anaconda.com/products/individual}}
Python distribution all necessary packages are already included.

\subsection{ana\_cont}
\noindent
The program files for the \verb|ana_cont| package can be downloaded from \anaconturl\
in several ways. It is however recommended to use git:
\begin{verbatim}
git clone https://github.com/josefkaufmann/ana_cont.git
\end{verbatim}
This will create a new directory \verb|ana_cont| in your current working directory.
Now change to this directory, \verb|cd ana_cont|. 
There all the code files are located. If you plan to do analytic continuations
by \pade\ interpolation, you have to compile the \pade\ core functions, which
are written in Cython\footnote{You can install it, e.g.~by \verb|pip install Cython|, 
if it is not already included in your Python distribution.}. 
The compilation is done by the setup script:
\begin{verbatim}
python setup.py build_ext --inplace
\end{verbatim}
Now, in a Python script, you have to insert the path to the \verb|ana_cont| package
into the module search path and import the package:
\begin{verbatim}
import sys
sys.path.insert(0, '/path/to/ana_cont/')
import ana_cont.continuation as cont
\end{verbatim}
Then the main classes, as described in \cref{sec:package}, can be accessed
as \verb|cont.AnalyticContinuationProblem| and \verb|cont.GreensFunction|.
Before writing a custom script it is recommended to go through some of our
learning resources on GitHub, e.g.
\begin{itemize}
	\item \verb|doc/basics.ipynb|\footnote{
	Files ending with '.ipynb' are jupyter notebooks. They can be displayed
	on github.com by clicking on them in a web browser. If the notebook is
	not displayed correctly, reloading the web page usually helps. 
	In order to execute
	the cells and make changes, it is necessary to install and use jupyter.},
\item \verb|doc/tutorial/tutorial.ipynb|,
\item \verb|scripts/example_fermionic.py|.
\end{itemize}
A tutorial related to the example case of \cref{fig:preblur-svo} is \verb|doc/tutorial_svo.ipynb|.

The Python scripts with GUIs, which are located in the \verb|scripts| directory,
are \verb|maxent.py| (fermionic maxent), \verb|maxent_bosonic.py| (bosonic maxent),
and \verb|pade.py| for the \pade\ method.
All details about their execution and usage are documented
in the wiki of our repository on GitHub at {\wikiurl}.
There we provide also tutorials with links for downloading files with test data.

\section{Conclusion}\label{sec:conclusion}
We have given an introduction to analytic continuation by means of \pade\ interpolation
and the maximum entropy method. These are implemented within our Python package
\texttt{ana\_cont}, which can be used with great flexibility for the analytic continuation of arbitrary bosonic and fermionic Green's functions. Extensions
to other analytic continuations outside the realm of quantum field theory are possible. 
Through casting
our established workflows in applications with GUIs, we have also made \maxent\
and \pade\ analytic continuation accessible for non-experts. Long-term users, on the other hand, will value the flexibility and adaptibility of the Python package and adjust the provided scripts to their individual needs.
\paragraph{Acknowledgment} We are indepted to Patrik Gunacker, who gave the initial incentive to the development of this code,
and to Andrey Katanin for sharing his knowledge and experience about the \pade\ method. 
Furthermore we thank Dominique Geffroy and Klaus Steiner, who motivated the implementation of many features
by their extensive usage of the code. For diligent testing and numerous valuable comments we are grateful to
Clemens Watzenböck, Jan Kune\v{s}, Oleg Janson, and Markus Wallerberger.
This work has been supported financially by the Austrian Science Fund (FWF) through projects  P30819 and P32044.


  \bibliographystyle{elsarticle-harv} 
  \bibliography{phd,add_refs}


%
%
%
\end{document}